\documentclass[%
 reprint,
superscriptaddress,
 amsmath,amssymb,
 aps,
]{revtex4-2}
\usepackage{xcolor}
\usepackage{graphicx}
\usepackage{dcolumn}
\usepackage{bm}

\begin{document}

\preprint{APS/123-QED}

\title{Towards electrical domain-wall control in polyacetylene-based electronic nano-devices}

\author{Leandro M. Arancibia}\thanks{These authors contributed equally to this work and are listed in alphabetical order.}
 \affiliation{%
Instituto Interdisciplinario de Ciencias Básicas (ICB-CONICET), Universidad Nacional de Cuyo,\\ Padre Jorge Contreras 1300, Mendoza 5502, Argentina
}

\author{Andrés I. Bertoni}\thanks{These authors contributed equally to this work and are listed in alphabetical order.}
 \affiliation{%
Instituto Interdisciplinario de Ciencias Básicas (ICB-CONICET), Universidad Nacional de Cuyo,\\ Padre Jorge Contreras 1300, Mendoza 5502, Argentina
}

\author{Cristián G. Sánchez}
 \affiliation{%
Instituto Interdisciplinario de Ciencias Básicas (ICB-CONICET), Universidad Nacional de Cuyo,\\ Padre Jorge Contreras 1300, Mendoza 5502, Argentina
}
\author{Alejandro M. Lobos}
 \email{Corresponding author: alejandro.martin.lobos@gmail.com}
 \affiliation{%
Instituto Interdisciplinario de Ciencias Básicas (ICB-CONICET), Universidad Nacional de Cuyo,\\ Padre Jorge Contreras 1300, Mendoza 5502, Argentina
}
\affiliation{%
Facultad de Ciencias Exactas y Naturales (FCEN), Universidad Nacional de Cuyo,\\ Padre Jorge Contreras 1300, Mendoza 5502, Argentina
}
\date{\today}

\begin{abstract}
We theoretically propose a polymer-based nano-device consisting of a single \textit{trans}-polyacetylene (tPA) molecule capacitively coupled to external gate voltages. We model the integrated device using a Su-Schrieffer-Heeger (SSH)-like Hamiltonian, and solve the self-consistent problem of coupled electronic and lattice degrees of freedom in the presence of the gate voltages. Interestingly, we demonstrate the emergence of multiple topological kinks or domain walls in the lattice distortion patterns, which can be externally controlled at the region of the gates. The local symmetry-breaking induced by the gates gives access to the different topological sectors of the SSH model characterized by a $\mathbb{Z}$  topological invariant, and allows  the emergence of quantized charges $Q_g=\pm ne$ at the gates, associated to the 
topological $n$-kink solutions. Given their topological origin, these solutions are remarkably stable as functions of the external gate voltage  $V_g$, except at specific values $V_g^{[n]}$ where topological transitions characterized by $\Delta n =\pm 1$ between different multi-kink solutions occur, and where the electronic subgap spectrum of the device is completely reconstructed in order to accomodate the extra charge. 
In practice, the device can be considered as a polymer-based organic quantum dot where charge quantization is inherently topological and robust, a fact that may be useful for potential technological applications.
\end{abstract}

\maketitle

\section{Introduction}
\label{sec:intro}
The emergent area of synthetic organic conductors based on polymers is an important field at the intersection of chemistry and physics,
of interest for both fundamental research and technological applications, such as organic electronics  ~\cite{Farchioni_Grosso_Organic_electronic_metals_book, Barford13_Electronic_and_Optical_Properties_of_Conjugated_Polymers}. Historically, \textit{trans}-polyacetylene (tPA), a linear chain of carbon atoms with alternating single and double bonds, was identified as the first organic conductor \cite{Heeger88_Solitons_in_conducting_polymers,Su79_Solitons_in_polyacetylene, Su80_Soliton_excitations_in_PA}. 
A key aspect of tPA is its double-degenerate ground state corresponding to the two possible ways to accommodate the Peierls dimerization pattern of single and double bonds. In its ground state, the system is an insulator characterized by a Peierls gap in the single-particle spectrum of excitations. However, tPA can be excited to a soliton-like state consisting of a topological kink or one-dimensional domain wall (to be used as synonyms in this work) in the lattice dimerization pattern, associated with a mid-gap electronic state and trapped excess charge \cite{Heeger88_Solitons_in_conducting_polymers,Su79_Solitons_in_polyacetylene, Su80_Soliton_excitations_in_PA}. The first experimental evidence of such soliton excitations in tPA was indirectly established via optical spectroscopy \cite{sethna1982photoinduced, blanchet1983photoexcitations} and magnetic electron paramagnetic resonance (EPR) experiments \cite{goldberg1979electron, weinberger1980electron}. However, these experimental techniques did not allow the direct access to individual solitonic excitations.

Recent progress in the synthesis of low-dimensional nanostructures via self-assembly have enabled to spatially resolve and identify one-dimensional solitons using scanning tunneling microscopy (STM) techniques \cite{cheon2015chiral, kim2017switching, lee2023mobile, park2022creation}. For instance, in the charge density wave system of indium atomic wires self-assembled on a silicon surface [i.e., In/Si(111) system], strong evidence of the presence of topologically protected chiral solitons has been found \cite{cheon2015chiral}.
Based on these findings, the idea of using solitons for information transport and to perform algebraic operations was proposed and demonstrated in Ref. \cite{kim2017switching}. The idea of using multiple solitons to encode discrete-information states was also recently introduced in Ref. \cite{lee2023mobile}. Other works recently proposed to use soliton excitations as the main mechanism for electronic transport in single-molecule nano-devices \cite{HernangomezPerez20_Solitonics_with_PA, park2022creation}.
In the case of tPA, on-surface nano-fabrication methods have demonstrated the ability to synthesize linear chains of tPA on clean metallic Cu(110) surfaces \cite{Wang19_Solitons_in_individual_PA_molecules}. In this study, the experimental signatures in the STM differential conductance $(dI/dV)$ indicated the presence of a trapped electronic state at a domain-wall induced at the  Cu/CuO interface, which can be interpreted as preliminary evidence for a non-chiral charged soliton state.

\begin{figure*}
    \centering
    \begin{minipage}[t]{0.35\textwidth}
        \raggedright
        \text{(a) } \vspace{1.5cm} \\
        \includegraphics[width=\textwidth]{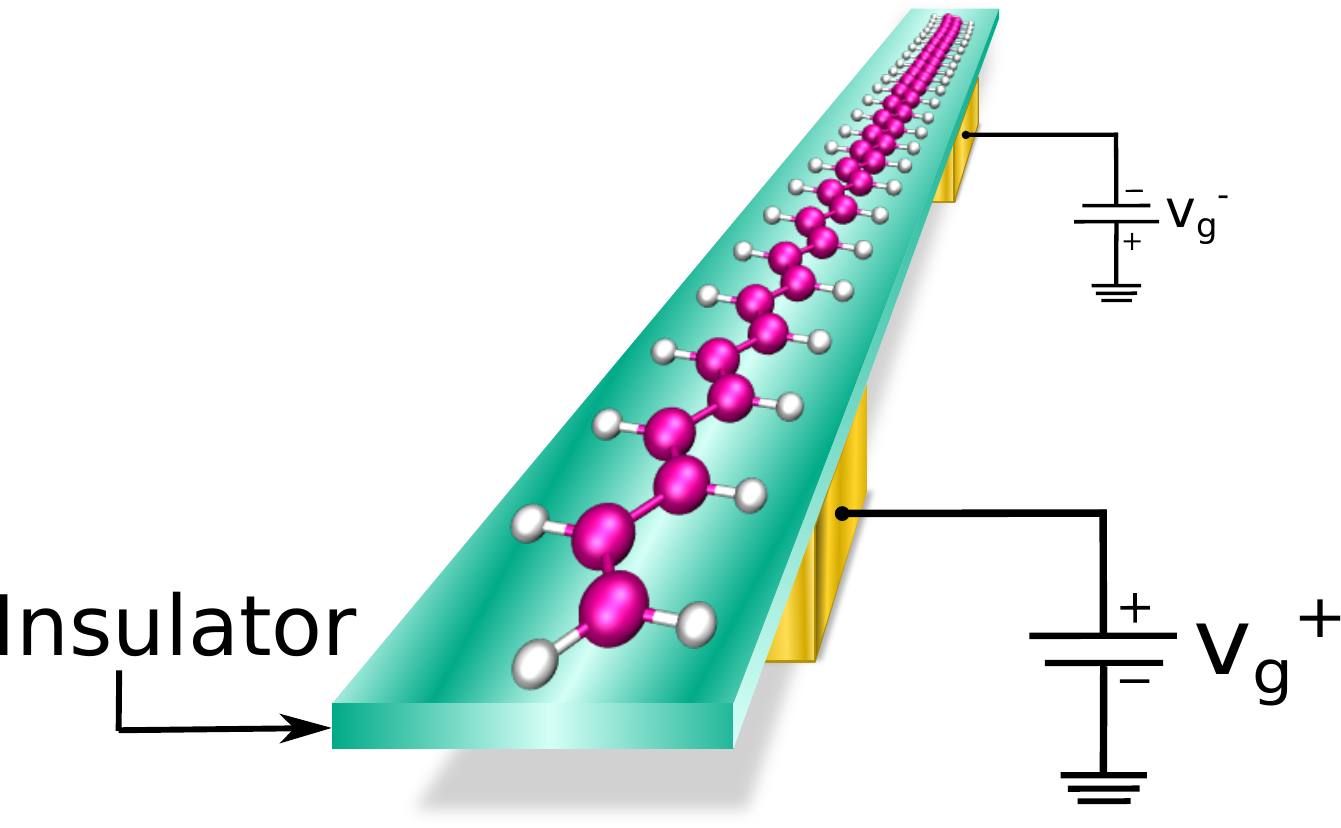}
    \end{minipage}
    \hfill
    \begin{minipage}[t]{0.6\textwidth}
        \raggedright
        \text{(b)} \\
        \includegraphics[width=\textwidth]{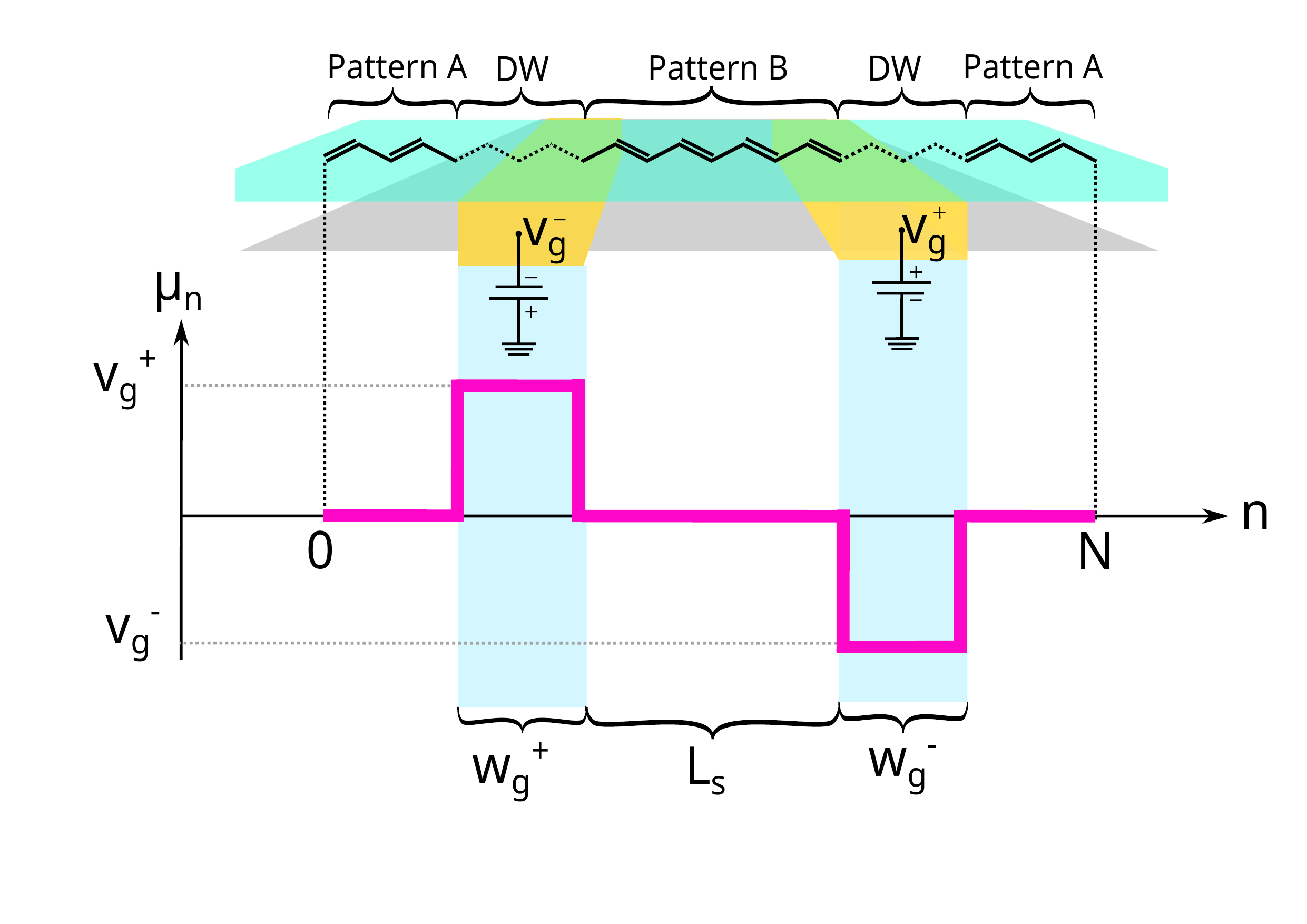}
    \end{minipage}
    \caption{(color online)  Scheme of the proposed tPA-based nano-device: a single finite trans-polyacetylene (tPA) molecule is deposited on an insulating thin film (e.g. $\text{SiO}_2$). The system is capacitively coupled to two independently adjustable gate voltages of equal width $w_{g}$ and separated by a distance $L_{s}$, that are connected to opposite voltages $V_{g}^{+}$ and $V_{g}^{-}$ (encoded in a phenomenological site-dependent chemical potential~$\mu_{n}$).}
    \label{fig:device}
\end{figure*}

Motivated by these recent advances and the prospects of novel electronic devices based on one-dimensional electronic solitons, in this work we theoretically study a linear tPA molecule deposited on an insulating film (e.g., $\text{SiO}_2$) and subjected to local gate voltages  $\pm V_g$, as shown in Fig. \ref{fig:device}. We modeled the system using a Su-Schrieffer-Heeger (SSH)-type model in which the gate voltages appear as a capacitive coupling term that modulates the chemical potential at the sites of the exposed molecular segments. Using a a self-consistent method to obtain the lattice configuration and the electronic structure of tPA in the ground state as a function of the applied gate voltage $V_g$, we demonstrate the ability to externally control the lattice dimerization pattern of the molecule by inducing domain-walls (DWs) or kinks at the region of the applied voltages. These  multi-kink configurations
 are remarkably stable as functions of the external gate voltage  $V_g$, except at specific values $V_g^{[n]}$ where topological transitions ocurr between solutions that differ in the number of topological kinks by $\Delta n=\pm 1$. At these transitions, the electronic subgap spectrum of the device is abruptly reconstructed in order to accomodate an extra quantized charge. 

From a practical perspective, the device would effectively behave as a molecular-scale quantum dot (QD). However, in contrast to standard semiconductor QDs for which charge quantization is based on the Coulomb blockade, here the quantization mechanism has an inherently topological origin, which results in superior robustness to disorder and other local perturbations over a wide range of parameters. 	As we show below, the proposed device holds promise for soliton manipulation technologies.

In Sec. \ref{sec:model} we present and discuss the details of the theoretical model; in Sec. \ref{sec:methodology} we describe the methodology and the numerical calculation approach to obtain the ground-state configuration of both the lattice and electron degrees of freedom. In Sec. \ref{sec:symmetry} we briefly discuss the symmetry and topological aspects of our modified SSH Hamiltonian and in Sec.
\ref{sec:results} we present our main results. Finally, we conclude in Sec. VI and provide details of some of our theoretical derivations in Appendix \ref{sec:appendix}. 

\section{Model}\label{sec:model}

Illustrated in Fig.~\ref{fig:device}, our model consists of a finite, linear tPA chain deposited on an insulating thin film (e.g. silica), with two gate voltages  placed underneath so that they are capacitively coupled to separate regions of the molecule. We assume that the gate regions are sufficiently separated from each other  so that crosstalk effects can be neglected.
These gates are connected to external voltage sources that are able to control the electron density of the exposed molecular segments by modifying their local electro-chemical potential. We model this device by the means of a SSH-like Hamiltonian \cite{Heeger88_Solitons_in_conducting_polymers, Su79_Solitons_in_polyacetylene}, modified to account for open-ended molecular chains \cite{Vos96_SSH_model_finite_length} and to describe the effect of the gate voltages. The SSH Hamiltonian is a celebrated model introduced to describe the interaction of $\pi$-conjugated electrons with the classical ionic degrees of freedom in linear polyalkenes. This rather simple model successfully accounts for many of the experimentally observed properties of tPA, such as the existence of a Peierls dimerization gap, and theoretically explains the emergence of solitons \cite{Farchioni_Grosso_Organic_electronic_metals_book, Heeger88_Solitons_in_conducting_polymers}. Since it realizes the simplest example of a topological insulator in one spatial dimension, the SSH model has become one of the most paradigmatic models in condensed-matter physics \cite{Asboth16_Short_course_on_TIs}. Explicitly, the model used in this work reads
\begin{align}
\hat{H} = \hat{H}_{\text{el}} + \hat{H}_{\text{latt}}\label{eq:H}, 
\end{align}
where the electronic Hamiltonian, $\hat{H}_{\text{el}}$, describes both the electron-lattice and the electron-gate interaction, and the lattice Hamiltonian, $\hat{H}_{\text{latt}}$, describes the classical coordinates of the (CH)$_n$ groups interacting via elastic terms:
\begin{align}
\hat{H}_{\text{el}} = & \sum_{n=1,s}^{N_{s}-1} \left[ -t_{0} + \alpha \left( u_{n+1} - u_{n} \right) \right] c_{n,s}^{\dagger} \, c_{n+1,s} + \text{H.c.}\nonumber \\ & + \sum_{n=1,s}^{N_{s}} \mu_n \, c_{n,s}^{\dagger} \, c_{n,s} \, , \label{eq:H_el} \\
\hat{H}_{\text{latt}} = & \sum_{n=1}^{N_{s}} \frac{p_{n}^{2}}{2M} + \sum_{n=1}^{N_{s}-1} \frac{K}{2} \left( u_{n+1} - u_{n} \right)^{2} \nonumber \\ & - \sum_{n=1}^{N_{s}-1} \Gamma \, (u_{n+1} - u_{n}) \, . \label{eq:H_latt}
\end{align}
The first term of Eq.~(\ref{eq:H_el}) describes the electrons in the $\pi$ bands, arising from the $2p_{z}$ orbitals of a tPA molecule. The molecule is  assumed to have a total of $N_{s}$ (CH) moieties, each one considered as a discrete ``site’’ in the chain. The operator $c_{n,s}$ ($c^{\dagger}_{n,s}$) annihilates (creates) an electron with spin $s$ at site $n$. The parameter ${t_{0} = 2.5 \, \text{eV}}$ is the hopping integral, ${\alpha = 4.1 \, \text{eV}/\text{\AA}}$ is the electron-lattice coupling parameter, and $u_{n}$ are the classical variables describing the deviation from equilibrium of the $n$-th CH group along the chain axis. The term in square brackets in Eq.~(\ref{eq:H_el}) results from the expansion of the full hopping integral up to first order in the small parameter $(u_{n+1} - u_{n} )$. The lattice parameter $a=1.22$  \AA 
\ is chosen as the unit of length in the problem in what follows. The above values of parameters  have been chosen to reproduce experiments and have been taken from Ref. \cite{ono1990motion}.

The last term of~$\hat{H}_{\text{el}}$ introduces the effect of the external gate voltages. To simplify the theoretical calculations, in this work we have utilized the microcanonical ensemble, where the total number of electrons $N_{\text{el}}$ is a conserved quantity. Additionally, we have introduced an overall half-filling condition, i.e. ${N_{\text{el}} = N_{s}}$, equivalent to global charge neutrality in the present context. This condition is ensured by constraining the electronic solutions in all the steps of the energy minimization process (see Sec. \ref{sec:methodology} below). Under the global half-filling constraint, the gate voltages produce a redistribution of the electronic charge throughout the system, favoring an excess of electron density at some sites at the expense of its depletion at other points in the chain. The microcanonical constraint effectively prohibits electron transfer to or from the environment, particularly the insulating film, which serves as a decoupling layer.

It is important to stress that our main purpose in this work is to extract the properties of a \textit{single} gate voltage. In that sense, the double-gate configuration is not essential and has to be considered as a theoretical trick that greatly simplifies the self-consistent calculations by keeping charge neutrality at all times in the calculation. Although the single-gate version of this model would be simpler to realize experimentally, it is more theoretically challenging. This is due to the need to equilibrate the electrochemical potential with an external charge reservoir, as is typical in open systems, making it necessary to adopt a grand-canonical formulation. By placing the gates sufficiently far away from each other, direct interaction between them is avoided, allowing for  the properties of a single-gate device to be inferred. As an additional simplifying assumption, we impose  exact anti-symmetric gates (i.e, identical to each other but with opposite voltages). As shown in Fig.~\ref{fig:device}(b), we define two distinct gate regions~$g^{+}$ and~$g^{-}$, each one spanning~$w_{g}$ sites associated with constant voltages $+V_{g}$ and $-V_{g}$, respectively, with ${V_{g} > 0}$. In Eq.~(\ref{eq:H_el}) these external gate voltages are included in a phenomenological site-dependent chemical potential~$\mu_{n}$, such that ${\mu_{n} = +eV_{g},~ n \in g^{+}}$ and ${\mu_{n} = -eV_{g},~ n \in g^{-}}$, being zero for non-gated sites (${\mu_{n} = 0 ,~ n \notin \{g^{+}, g^{-}\}}$). The anti-symmetry of the external voltages  is a ficticious symmetry that simplifies the extrapolation of  our results to the case of a single gate in open conditions, since it ensures particle-hole symmetry in the electronic spectrum and charge neutrality, and therefore the emergence of quantized charges $+e$ and $-e$ occurs simultaneously at both gates as a function of $V_g$. For this work, we have considered a tPA molecule with ${N_{s} = 200}$ sites,  and anti-symmetric  gate voltages of width ${w_{g} = 40 \, a \approx 5 \, \text{nm}}$, placed at distance of ${L_{s} = 80 \, a \approx 10 \, \text{nm}}$ from each other. Note that these values of $w_g$ and $L_s$ are within experimental reach using, e.g., electron-beam lithography techniques. 

The first term in Eq.~(\ref{eq:H_latt}) provides a classical description for the lattice ions, where ${M = 2.152} \, \times 10^{-26}$ kg is the mass of the $\text{(CH)}_{n}$ groups and ${K = 21 \, \text{eV}/\text{\AA}^{2}}$ is the effective spring force constant of the C-C bonds. As was already noted by Su and Schrieffer ~\cite{Su80_Dynamics_of_solitons_in_PA}, and by Vanderbilt and Mele ~\cite{Vanderbilt80_Disorder_in_tPA}, the self-consistent treatment of the open-boundary SSH model is technically complicated by a spurious tendency of the chain to shrink under the effect of the unconstrained electronic linear coupling to the $\pi-$electrons. This is of course an artifact of the linear approximation in the hopping integral combined with the harmonic approximation,  which is absent in simplified (and more usual) versions of the SSH model where self-consistency is avoided, and where the lattice degrees of freedom are  externally provided via an alternating hopping $t_0+\left(-1\right)^n\delta t$.
Of course, this spurious shrinking could be avoided by using more sophisticated treatments (e.g., using proper \textit{ab initio} potentials), but this would make  our numerical computations very costly. Thus, in order to compensate for this undesired effect while keeping the simplicity of the SSH model, we followed Refs.~\cite{Su80_Dynamics_of_solitons_in_PA, Vanderbilt80_Disorder_in_tPA, Vos96_SSH_model_finite_length} and introduced an additional term, ${\sum_{n=1}^{N_{s}-1} \Gamma \, (u_{n+1} - u_{n}) = \Gamma \, (u_{N_{s}} - u_{1})}$, representing  an effective ``stretching’’ force applied to the ends of the chain to compensate for the spurious contraction. We have chosen ${\Gamma = 5.15 \, \text{eV}/\text{\AA}}$ to reproduce the same bulk properties seen in experiments (i.e., values of the Peierls distortion, gap size, etc.). Since the stretching term only introduces distortions that are highly localized at the ends of the chain, it does not affect our most important results which concern the effect of localized defects at the region of the gates, sufficiently away from the edges.

Previous experimental works using chemical doping techniques reported evidence of charged intra-gap states associated with solitonic or polaronic excitations appearing at random positions along tPA chains \cite{Heeger88_Solitons_in_conducting_polymers, Chiang77}. In that sense, the gating mechanism proposed here can be considered as a local ``electronic doping", with the difference of being site-specific and fully controllable, thus enabling the functionalization of the device. Moreover, unlike simplified versions of the SSH model where the dimerization pattern is externally imposed, we have solved the self-consistent equations for the coupled lattice and electronic degrees of freedom for a wide range of values of $V_g$ (see Sec. \ref{sec:methodology}). This point is crucial for the emergence of topological phase transitions controlled through the external gate voltages, which act as experimental control knobs over the lattice configuration $\{u_n\}$ at the specific regions of the gates. Interestingly, these voltage-controlled transitions are associated with a change in the number of topologically-protected quantized charges at these regions (see Section~\ref{sec:results}). Finally, we note that the presence of the gates locally breaks particle-hole and/or inversion symmetry, and allows to explore the full $\mathbb{Z}$ topology of the SSH Hamiltonian \cite{Ryu10_Topological_classification, Kitaev_TI_classification}. This is in contrast to the case of the translationally-invariant SSH model with particle-hole, time-reversal, and inversion symmetries,
where only two phases (either topologically trivial or non-trivial) can be achieved,  effectively reducing the topological invariant to $\mathbb{Z}_2$ \cite{Asboth16_Short_course_on_TIs}.  

\section{Methodology}\label{sec:methodology}

In order to solve the coupled electron-lattice problem at temperature  $T=0$, we employ the Hellmann-Feynman theorem \cite{feynman39}:
\begin{equation}\label{eq:Helmann-Feynman}
\frac{\partial E_{\mathrm{GS}}}{\partial R_{n}} = \left\langle \Psi_{\mathrm{GS}} \left| \frac{\partial \hat{H}}{\partial R_{n}} \right| \Psi_{\mathrm{GS}} \right\rangle = 0 \, ,
\end{equation}
where $|\Psi_{\mathrm{GS}}\rangle$ is the electronic ground-state wavefunction, $E_{\mathrm{GS}}$ is the total ground-state energy (i.e., ${\hat{H} \, |\Psi_{\mathrm{GS}}\rangle = E_{\mathrm{GS}} \, |\Psi_{\mathrm{GS}}\rangle}$), and ${R_{n} = na + u_{n}}$ is the position of the $n-$th site along the chain axis. Eq.~(\ref{eq:Helmann-Feynman}) leads to the following system of coupled equations:
\begin{widetext}

\begin{align}
K \, \left( R_{2} - R_{1} - a \right) - \Gamma + \alpha \sum_{s}\sum_{\nu=1}^{\text{occ}} \left[ \psi_{\nu,s}^{*}(2) \, \psi_{\nu,s}(1) + \text{H.c.} \right] = & \, 0 \, , \label{eq:HF_L} \\
K \, \left( R_{n{+}1} - 2 R_{n} + R_{n{-}1} \right) - \alpha \sum_{s}\sum_{\nu=1}^{\text{occ}} \left[ \left( \psi_{\nu,s}^{*}(n{+}1) - \psi_{\nu,s}^{*}(n{-}1) \right) \, \psi_{\nu,s}(n) + \text{H.c.} \right] = &  0,  \ \left( \forall  \  n  \text{ such that } 1<n<N_{s}\right), \label{eq:HF_mid}\\ 
-K \, \left( R_{N_{s}} - R_{N_{s}{-}1} - a \right) + \Gamma - \alpha \sum_{s}\sum_{\nu=1}^{\text{occ}} \left[ \psi_{\nu,s}^{*}(N_{s}{-}1) \, \psi_{\nu,s}(N_{s}) + \text{H.c.} \right] = & \, 0 \, . \label{eq:HF_R}
\end{align}
\end{widetext}

In the above equations, $\psi_{\nu,s}\left(n\right)$ is the probability amplitude of the $\nu$-th eigenstate at the $n$-th site, i.e. ${|\psi_{\nu,s}\rangle \equiv \sum_{n=1}^{N} \psi_{\nu,s}\left(n\right) c^\dagger_{n,s} |0\rangle}$, which satisfies the single-particle eigenvalue equation:
\begin{align}\label{eq:eigenvalue}
\hat{H}_{\text{el}} \, |\psi_{\nu,s} \rangle = \epsilon_{\nu} \, |\psi_{\nu,s} \rangle ,
\end{align}
with $\epsilon_{\nu}$ as the corresponding single-particle energy. Moreover, note that summations run over the doubly occupied (``occ’’) one-electron states since $N_{\text{el}}^{\uparrow} = N_{\text{el}}^{\downarrow}$ due to the SU(2) symmetry of the model. Note also that Eq.~(\ref{eq:eigenvalue}) is solved for a given specific configuration of the lattice deformation field $R_{n}$; therefore, the set of Eqs.~(\ref{eq:HF_L})-(\ref{eq:eigenvalue}) forms a complete set of coupled non-linear equations that must be solved self-consistently. As we will see in Sec. \ref{sec:results} below, this self-consistency is crucial to understand the emergence of robust multi-kink configurations hosting quantized charges at the gates, and the occurrence of topological phase transitions as the parameter $V_g$ is varied. Finally, note that for our model of non-interacting electrons, the electronic ground state is simply written as  ${\left| \Psi_{\mathrm{GS}} \right\rangle = \prod_s \prod_{\nu=1}^{N_{\text{el}}} c_{\nu,s}^\dagger |0\rangle }$, with the fermionic operators $c_{\nu,s}$ defined as ${c_{\nu,s} = \sum_{n=1}^{N_s} \psi_{\nu,s}\left(n\right) \, c^\dagger_{n,s} }$.

To obtain the electron-lattice solution of minimum energy, we have implemented an iterative numerical routine  that starts with a random distribution of ion positions $R_{n}^{(0)}$. This distribution is used to build the Hamiltonian $\hat{H}_{\text{el}}$, from where the electronic eigenstates $|\psi_{\nu,s} \rangle$ are obtained via Eq. (\ref{eq:eigenvalue}). These eigenstates  are then used to calculate the new ion positions, $R^{(1)}_{n}$, by solving the linear system of Eqs.~(\ref{eq:HF_L})-(\ref{eq:HF_R}). We iterate this process until a convergence criterion is met, namely when the root-mean-square deviation (RMSD) of the ion positions with respect to the preceding iteration falls below $10^{-8} \, a$. It is important to highlight that at each step of the optimization routine we ensured that the total number of electrons $N_\text{el}$ is kept  constant (hence keeping the system neutral), as this corresponds to the microcanonical ensemble under the half-filling condition.

\section{Symmetry and topological aspects}\label{sec:symmetry}
The symmetry properties of the Hamiltonian Eq. (\ref{eq:H}) determine its topological class and its topological phases \cite{Ryu10_Topological_classification, Kitaev_TI_classification}. In the first place, we note that under the inversion with respect to the middle point of the molecule, i.e., $n \to N_s+1-n $, the Hamiltonian Eq. (\ref{eq:H}) has the following symmetry property:
\begin{align}
\hat{H}(\{u_n\},V_g)&\to
\hat{H}(\{-u_n\},-V_g).\label{eq:inversion}
\end{align}
To show this, we have used the aforementioned anti-symmetry of the gate voltages.  For an externally supplied dimerization pattern  $u_n=(-1)^n u_0$ (in which case self-consistency can be avoided and we can set $\Gamma=0$), and for $V_g=0$, we immediately note that inversion symmetry implies that
the values $\pm u_{0}$ are degenerate. In this context, it is customary to define the \textit{staggered} lattice distortion as:
\begin{align}\label{eq:staggered_field}
\bar{y}_{n} \equiv & \, (-1)^{n} \left[ u_{n+1} - u_{n} \right], \nonumber\\
= & \, (-1)^{n} \left[ R_{n+1} - R_{n} - a \right]\, ,
\end{align}
where the fast oscillations have been removed via the factor $(-1)^{n}$. As a result, $\bar{y}_{n}$ is a smooth variable on the scale of the lattice parameter $a$, allowing to identify  regions with a specific dimerization pattern. In this way, the presence of a topological kink in the system can be associated to the points where $\bar{y}_{n}$ crosses zero. Consequently, a region with a constant staggered field of ${\bar{y}_{n} = \pm 2 u_{0}}$ corresponds to one of the two possible Peierls-dimerization patterns of the tPA molecule obtained from the minimization of the total energy. Either of these two uniform solutions leads to a Peierls gap ${\Delta_{0} = 4 \alpha \left| u_{0} \right|}$ between the filled valence band and the empty conduction band, at ${T = 0}$. For the parameters used in the present work, we have determined the Peierls gap at ${V_{g} = 0}$ to be ${\Delta_{0} = 0.628 \, \text{eV}}$; this value will be used as a reference in what follows. In the absence of external voltages, we have obtained a uniform profile with ${|\bar{y}_{n}| = 0.0755 \, \text{\AA}}$
in the range of voltages up to ${\sim 0.4 \, \text{V}}$ (see Fig.~\ref{fig:DWs}(a)).

In addition, under the particle-hole transformation $c_{n,s}\to c^\dagger_{n,s}$ the electronic Hamiltonian Eq. (\ref{eq:H_el}) transforms as
\begin{align}
\hat{H}_\text{el}(\{u_n\},V_g)&\to-\hat{H}_\text{el}(\{u_n\},V_g).\label{eq:particle_hole}
\end{align}
Again, the anti-symmetry of the gate voltages has been used here. However, even though $\hat{H}_\text{el}$ has the above \textit{global} particle-hole symmetry, it is important to note that this symmetry is still \textit{locally broken} by the gate voltages. This local breaking of particle-hole symmetry is crucial for understanding our results and allows for a full exploration of the 
$\mathbb{Z}$ topology of the SSH model. Note that this cannot be achieved  in more constrained situations where particle-hole symmetry is locally enforced at every site in the chain, resulting in more restricted  configurations. Indeed, while  the existence of regions where both  inversion and particle-hole  symmetries are preserved (i.e., not broken by the presence of the gates) is crucial to give definite topological properties to the Hamiltonian Eq. (\ref{eq:H}), the local enforcement of these symmetries \textit{at all sites} become too restrictive and only two topologically distinct ground states can be  realized (either topologically trivial or non-trivial). We mention in passing that while in translationally-symmetric one-dimensional models with particle-hole, time-reversal and inversion symmetries, a topological insulator characterized by a $\mathbb{Z}$ topological invariant can be achieved, as is the case of the BDI or AIII classes in the Cartan classification \cite{Ryu10_Topological_classification, Kitaev_TI_classification}; in practice, only two topologically distinct ground states  can be  realized, reducing this invariant to $\mathbb{Z}_2$. Here, by locally breaking these symmetries with the gate voltages (and considering them sufficiently separated), the full $\mathbb{Z}$ topology of the SSH model can be explored.

Finally, we mention that the emergence of  topological  kinks in the SSH Hamiltonian can be modeled in the continuum limit ${a \rightarrow 0, } \, { N_{s} \rightarrow \infty }$, by the means of the massive Dirac equation, and this problem has been addressed in the seminal work by Jackiw and Rebbi  \cite{Jackiw76_Jackiw_Rebbi_soliton}, and also by Takayama et al \cite{Takayama80_Continuum_model_for_PA}. In the continuum model,  the distortion profile ${\bar{y}_{n} \rightarrow \Delta(x)}$, with ${x = na}$, can be interpreted as an effective sign-changing ``mass’’ term. For an infinite system with a kink centered at $x=0$, this mass term adopts the simple functional form ${\Delta(x) = 2 u_{0} \tanh (x / \xi )}$, where ${\xi = \hbar v_{F} / \Delta_{0}}$ is the width of the kink (with $v_{F}$ being the Fermi velocity of the massless Dirac fermions\cite{Heeger88_Solitons_in_conducting_polymers}). Therefore, $\Delta(x)$ is a solution that interpolates between the constant values $\pm 2 u_{0}$ corresponding to single Peierls dimerization patterns at  ${x \rightarrow \pm \infty}$. This model predicts topological intra-gap electronic states that are localized at the kink within a distance $\xi$ and, in the case of local enforcement of particle-hole symmetry, they emerge at the Fermi energy  ${\epsilon = 0}$. These bound states are able to accommodate an extra charge and/or spin, giving rise to the celebrated fractionalized solitonic excitations in tPA \cite{Su80_Soliton_excitations_in_PA, Heeger88_Solitons_in_conducting_polymers}.

\section{Results}\label{sec:results}

\begin{figure*}[t]
    \centering
    \includegraphics[scale=0.85]{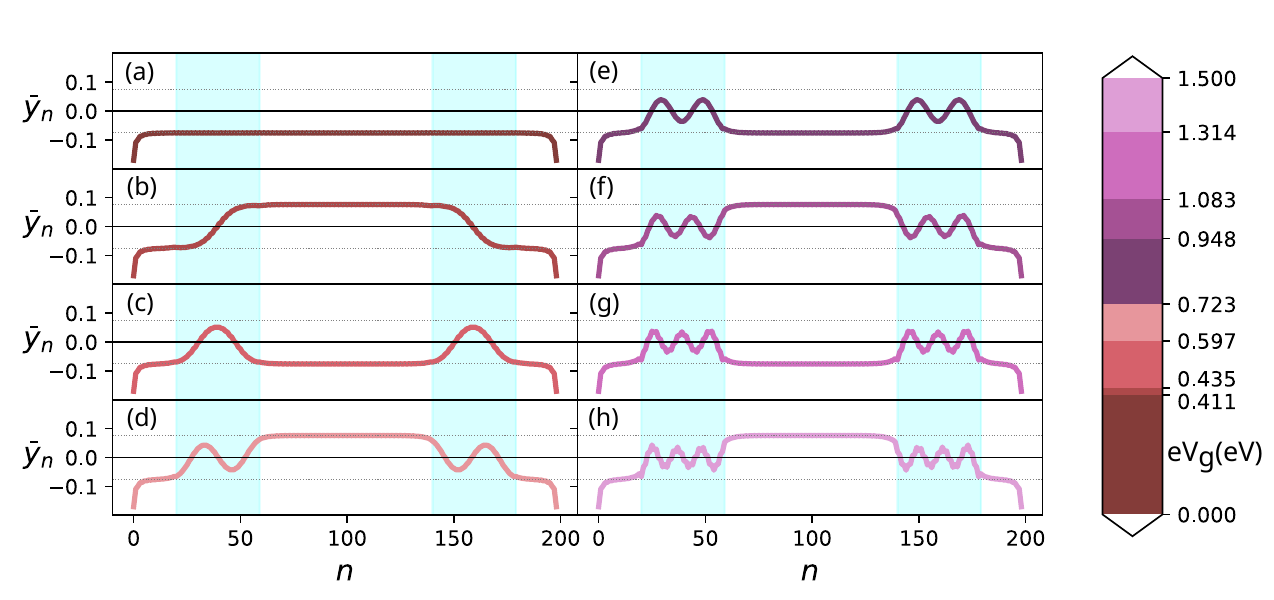}
    \caption{(color online) Staggered (lattice) distortion profiles $\bar{y}_{n}$ for the distinct lowest-energy configurations (i.e., global minima) that occur at applied voltages $V_{g}$ up to $1.5 \, \text{eV}$. The segments with a constant value of $\bar{y}_{n}$ correspond to one of the two possible Peierls dimerization patterns of tPA. The color-shaded regions indicate the location of the gate voltages, meaning that these sites can have ${\mu_{n} \ne 0}$. In the gate regions, the points where $\bar{y}_{n}$ crosses zero correspond to the center of DWs. Note that the number of DWs per gate changes in an abrupt and discrete way only at critical values of $V_{g}$, whereas all solutions within each voltage range of stability (color-coded according to the color bar on the right) have their profiles visibly unaffected. While in each stability range they appear to be a single curve, we stress that there are actually multiple nearly identical curves.
     }
    \label{fig:DWs}
\end{figure*}

In Fig. \ref{fig:DWs} we show our main results for the lattice equilibrium configuration as a function of the parameter $V_g$. In the interval $V_g \in [0,0.411]$ V (see Fig. \ref{fig:DWs}(a)) we obtain a uniform configuration stabilized at $\bar{y}_n = - 0.0755$ \AA, 
representing one of the two possible dimerization patterns in the unperturbed SSH Hamiltonian (pattern A in Fig. \ref{fig:device}(b)). We stress that although the plot in Fig. \ref{fig:DWs}(a) appears to be a single curve, \textit{it actually corresponds to multiple nearly identical curves} obtained for different values of $V_g$ within the above interval. This shows that this uniform configuration is remarkably stable to small or even moderate values of $V_g$. On a side note, the  localized downturns near the edges are distortions
introduced by the stretching $\Gamma$ term and should be disregarded in what follows, as they hold no physical
significance at the region of the gates, shown as shaded cyan areas in Fig. \ref{fig:DWs}. As the external voltage $V_{g}$ is increased beyond a critical value of $V_{g}^{[1]} \approx 0.411 \, \text{V}$, the system undergoes an abrupt transition to a different lattice distortion profile containing a single topological kink per gate (see Fig.~\ref{fig:DWs}(b)). 
Since these two configurations belong to different topological sectors (i.e., they differ in the number of kinks), they cannot be smoothly connected and the system must resolve this situation via a topological phase transition. Interestingly, the value of $V_{g}^{[1]}$ can be independently estimated by equating the energy needed to create a soliton in a non-gated system (relative to the uniformly-dimerized ground state), i.e.,  $\Delta E_S= 2 \Delta_0 /\pi$ \cite{Barford13_Electronic_and_Optical_Properties_of_Conjugated_Polymers,Heeger88_Solitons_in_conducting_polymers},  to the electrostatic energy $e V_g$. From the resulting equation $eV_g^{[1]}=2 \Delta_0 /\pi$,  we obtain $eV_g^{[1]}\simeq 0.400$ eV, very close to the numerically obtained critical value
 $eV_g^{[1]}\simeq 0.411$ eV. This suprisingly accurate estimation indirectly serves as a sanity check of our calculations and of our hypothesis of sufficiently separated gates (i.e., since there are no residual interactions between the gates at $g^+$ and $g^-$).  

In terms of the continuum model discussed in the previous Sec. \ref{sec:symmetry}, the distortion profile of Fig.~\ref{fig:DWs}(b), which contains one kink per gate separated by a distance ${L_{s} \gg \xi}$, can be reasonably reproduced with the ansatz:
\begin{eqnarray}
\Delta (x) \approx -2 u_{0} \tanh\left( \frac{x - x_{g^{+}}}{\xi} \right) . \tanh \left( \frac{x - x_{g^{-}}}{\xi} \right), 
\end{eqnarray}
where $x_{g^{+}}$ and $x_{g^{-}}$ refer to the center of the respective gates. In other words, perturbing the tPA molecule via an external voltage favors the accumulation of excess charge bound to the gate regions, via the generation of charged kinks. In this context, it should be noted that the separation between the edges and the gate voltages  is also much larger than the characteristic length $\xi$ and, as mentioned above, the lattice distortion at the edges does not affect our results on the gated regions. For the case at hand (see Fig.~\ref{fig:DWs}(b)), a positive charge ${Q_{g}^{+} = +e}$ becomes localized at the DW on $g^{+}$ and, because of the anti-symmetry of the external gate voltage  profile, a compensating negatively charged DW  simultaneously emerges at $g^{-}$ with charge ${Q_{g}^{-} = -e}$, maintaining the overall charge neutrality. Interestingly, in contrast to the non-gated SSH model where the particle-hole symmetry is enforced at all sites, here it was not possible to identify an individual one-electron (bound) intra-gap state responsible for the excess charge. Instead, each localized charge emerges as a collective phenomenon involving all the occuppied states in the valence band.

\begin{figure}
    \centering
    \includegraphics[scale=0.47]{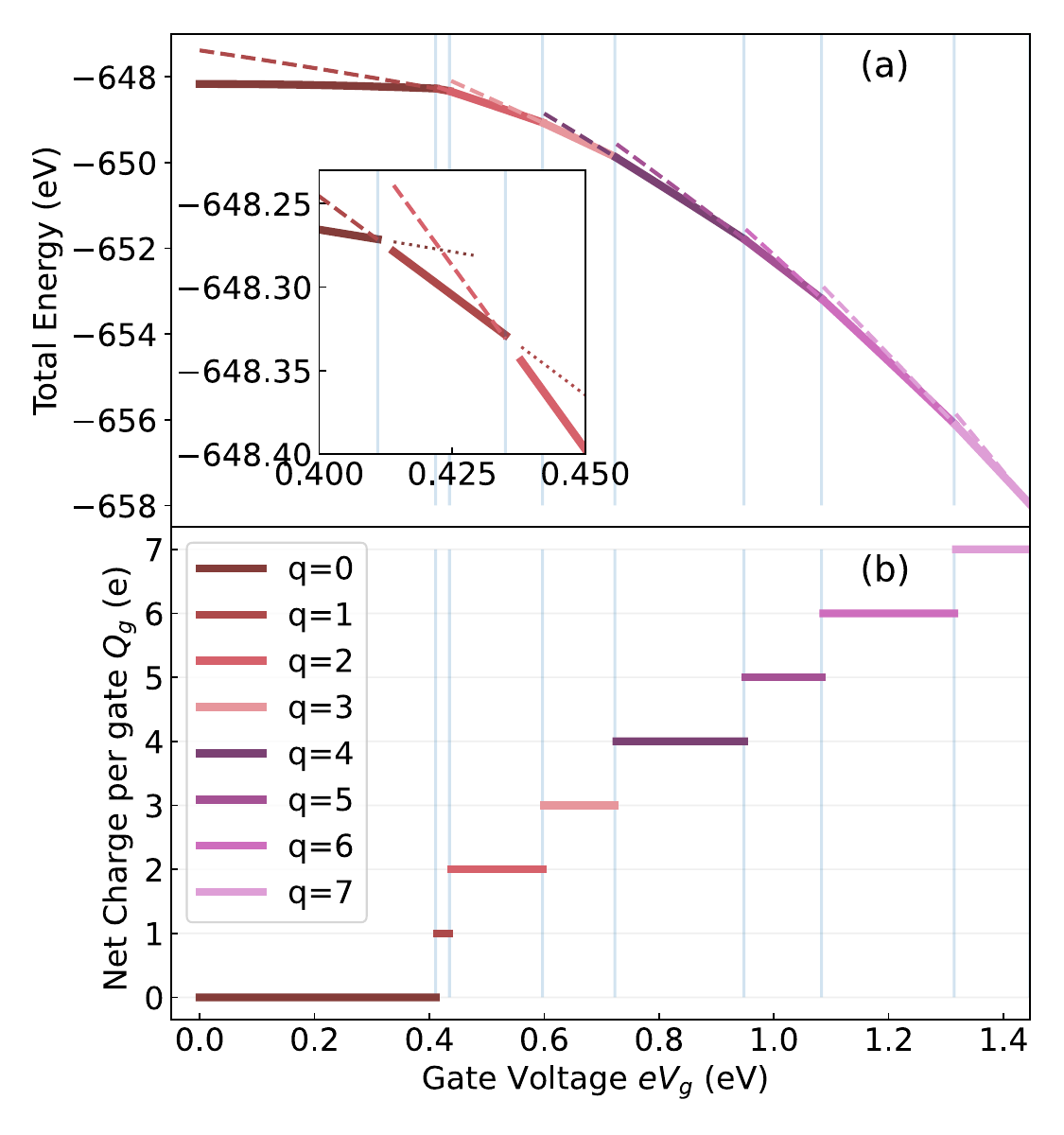}
    \caption{(color online) Total energy of the system (a) and absolute net charge per gate (b) as functions of applied voltage $V_{g}$. In both panels we use the same color coding as in Fig.~(\ref{fig:DWs}) to refer to the lattice configurations with a given number of DWs per gate. In panel (a), the solid lines correspond to the lowest-energy the system can be at each $V_{g}$. The dashed lines correspond to low-lying excited molecular states; since these are local minima, these states can be reached by starting from initial ion positions close to the desired non-optimal configuration. Notice how the quantum phase transitions occur at critical voltages, i.e. where there is a crossing of energy levels. In panel (b) we show the number of accumulated charges per gate from the CDWs present in the lowest-energy configuration for each $V_{g}$. To compute these values, we integrated the excess charge per site (see Section~A of the supplementary material) for a gate region and took its absolute value, resulting in an equal value of charge for both gates. In practice, a small number of sites immediately outside the gates had to be included in the integration in order to cancel out a very small charge contribution coming from spurious dipoles appearing at the sharp discontinuities in $V_{g}$. These artifacts have no physical significance for the modeled system and dissipate as the perturbation profile becomes smoother (not shown).
}
\label{fig:energy_and_charge}
\end{figure}

By further increasing $V_g$, other topological transitions occur at values ${V_{g}^{[2]}, \, V_{g}^{[3]}, \, \dots}$ where new self-consistent solutions with an increasing number of kinks per gate become stabilized, as shown in Fig.~\ref{fig:DWs}(c-h). 
For each of these new $n$-kink configurations, we can define a different voltage range  where they are robust and stable. Note that all these configurations obey the inversion symmetry mentioned above with respect to the middle-point of the molecule obtained in Eq. (\ref{eq:inversion}). 
Moreover, a  configuration with $n$ kinks per gate hosts a total quantized charge $Q_g=\pm ne$ localized at that gate. From here, it is simple to see that at each topological transition the total charge must abruptly increase (or decrease) in units of $e$. To illustrate this point, in Figs.~\ref{fig:energy_and_charge}(a) and (b) we show, respectively, the total energy of the system (i.e., electronic plus lattice) and the absolute net charge contained at each gate region, as functions of the applied voltage $V_{g}$. In panel (a) we show the evolution of the ground-state energy (solid thick lines), along with the energy of the first excited configurations (thinner dashed lines), as functions of $V_g$. Analogously to the case of Fig.~\ref{fig:DWs}, we emphasize that a new self-consistent calculation has been performed for each value of $V_g$. We note that, since these configurations are local minima of the global energy, a local stability region can be defined. Therefore, even if they are not the ground-state solution, they can still exist as stable excited states. In this Figure, the occurrence of topological phase transitions can be clearly identified with the level crossings between configurations having a different number of kinks. These topological phase transitions are therefore first-order phase transitions.

In panel (b) we show  the total charge $Q_g$ bound at each gate region, which follows a perfect quantization step corresponding directly to the stability range of a given configuration. 
This implies that the voltage-induced quantized charges are as stable to variations in $V_g$ as the lattice configurations themselves. There is a simple analytical connection between the two panels of Fig.~\ref{fig:energy_and_charge}, as the excess charge per gate $Q_{g}$ can be obtained by deriving the total ground-state energy $E_{\text{GS}}$ with respect to the external potential $V_{g}$, making use of the Hellman-Feynman theorem:
\begin{align}
\frac{\partial E_{\mathrm{GS}}}{\partial V_{g}} = & \left\langle \Psi_{\mathrm{GS}} \left| \frac{\partial \hat{H}_{\text{el}}}{\partial V_{g}} \right| \Psi_{\mathrm{GS}} \right\rangle \, , \nonumber\\
= & -2 \, Q_{g},\label{eq:localized_charge}
\end{align}
where the factor 2 is obtained due to the anti-symmetry of the gates, as shown in Appendix \ref{sec:appendix}. The remarkable simplicity of this expression is related to the fact that the capacitive coupling term contributes to the total energy as a electrostatic potential energy term due to the charges accumulated (per gate) by the effect of $V_g$.


\begin{figure}[t]
    \centering
    \includegraphics[scale=0.45]{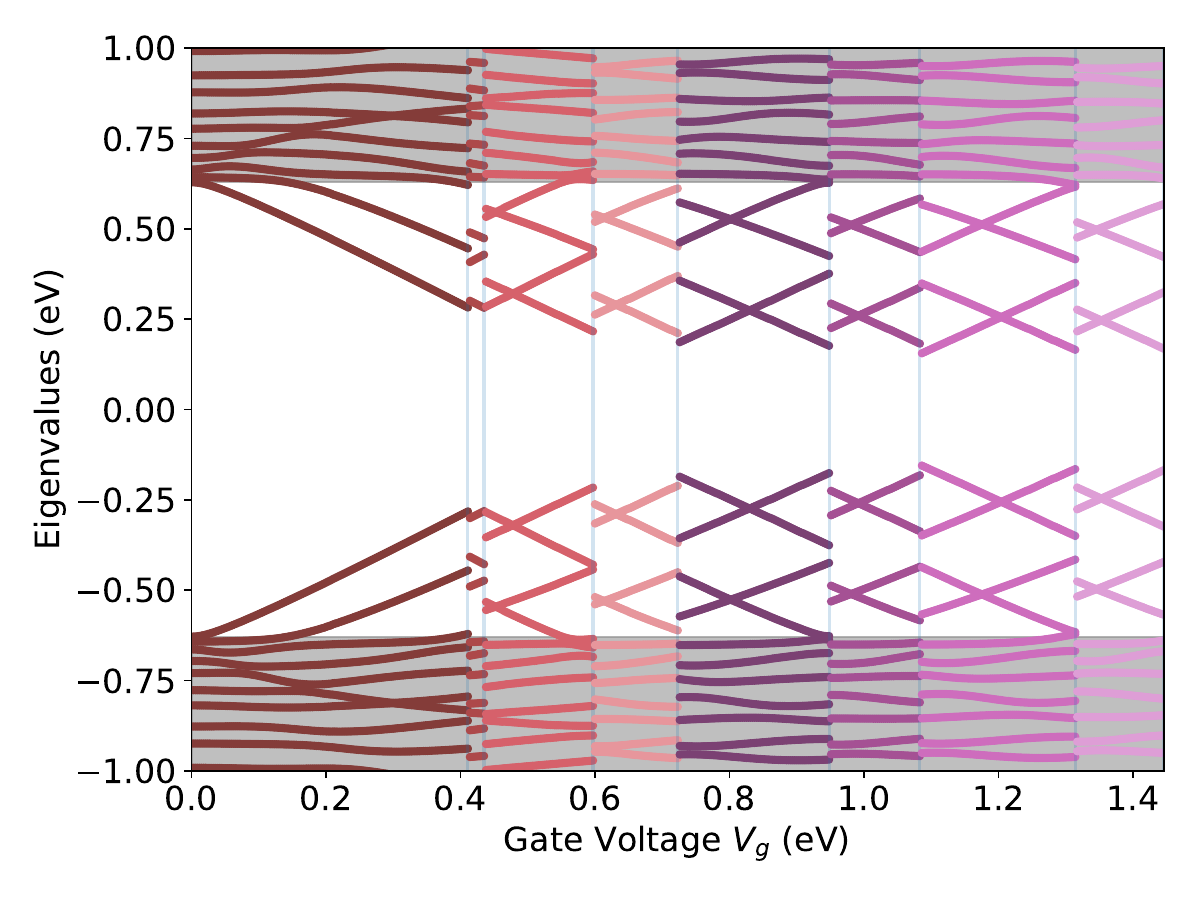}
    \caption{(color online) Electronic single-particle energy spectrum in the sub-gap region (i.e., white area)  as a function of the applied voltage $V_{g}$. The gray-shaded areas correspond to the valence- and conduction-band continuum.  The different colors denote the different ranges of voltage stability (here the same color code as in Fig.~\ref{fig:DWs} is used) and the different topological sectors. Note that the electronic states vary continuously with $V_{g}$ without affecting the stability of the lattice multikink configuration (see  Fig.~\ref{fig:DWs}). The occurrence of  topological phase transitions appears as abrupt discontinuities in the evolution of the spectrum with respect to  $V_{g}$.}
    \label{fig:eigenenergies}
\end{figure}

In Fig.~\ref{fig:eigenenergies} we show the evolution of the electronic spectrum inside the energy gap, as a function of $V_{g}$. Here, the topological phase transitions can be seen as abrupt changes in the electronic single-particle spectrum, with changes being more pronounced for the intra-gap states. Note that while the electronic spectrum  varies smoothly within intervals of phase stability, the lattice distortion profile remains robust and visibly unaffected. It is also worth noting that the spectrum has particle-hole symmetry, as dictated by Eq. (\ref{eq:particle_hole}).

In contrast to the single-kink case in the Jackiw-Rebbi soliton problem  \cite{Jackiw76_Jackiw_Rebbi_soliton}, where the extra charge can be associated to a particular state in the single-particle spectrum (i.e., specifically, the $\epsilon=0$ state), such an association is not possible here. Instead, charge quantization emerges as a collective phenomenon, where all occuppied states are involved. At the core of this phenomenon is the fact that the gate voltages locally break the particle-hole symmetry of the single-particle solutions, causing the zero-energy states of the otherwise unperturbed SSH model (or the Jackiw-Rebbi problem) to shift to lower or higher energies, depending on where these states have most of their weight. Phenomenologically, we can explain this as a non-trivial effect of the local gating combined with scattering against the multiple-kink lattice configuration, which generates a non-trivial phase shift in all the electronic states and the binding of charges via the Friedel's sum rule \cite{mahan}.

\begin{figure}
    \centering
    \includegraphics[scale=0.55]{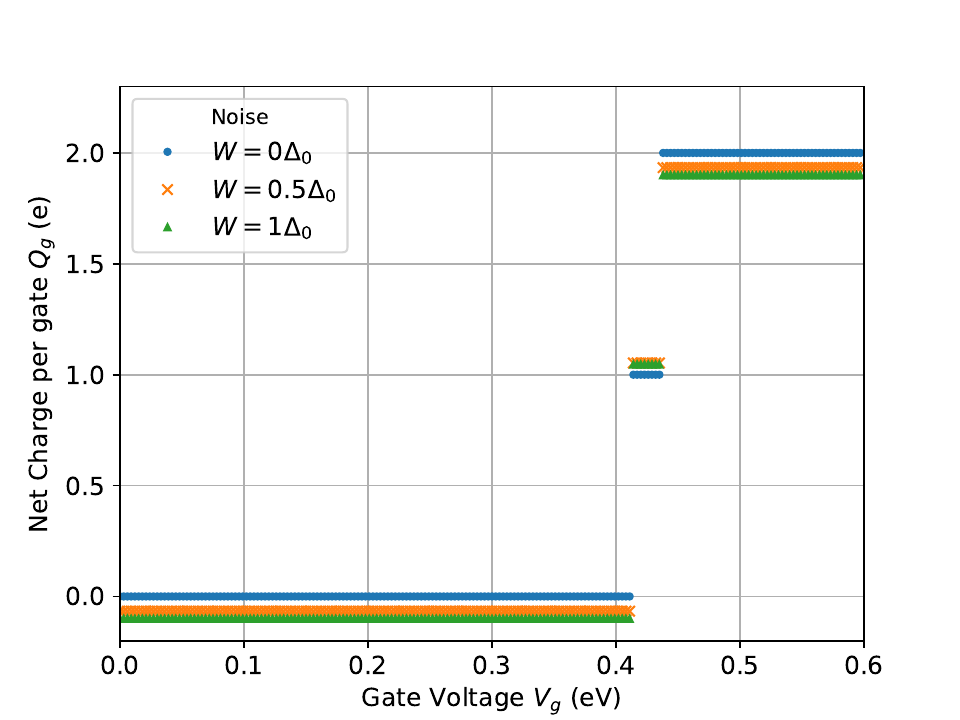}
    \caption{(color online) Net charge localized per gate $Q_g$ as a function of the external gate voltage $V_g$, for different realizations of the random on-site potential $W$ express in terms of the gap $\Delta_0$. Net charge quantization per gate voltage is lost as noise increases.}
\label{fig:disorder}
\end{figure}

Finally, we briefly discuss the effect of lifting the protecting symmetries.  
While the local breaking of particle-hole and inversion symmetry is essential for exploring the topological phase diagram in this problem, the presence of protecting symmetries in the non-gated regions is equally crucial for the existence of well-defined topological quantities.
To test and quantify the robustness of the device against the presence of symmetry-breaking perturbations, we have studied the effect of adding an on-site disorder potential term
\begin{align}
\hat{H}_\text{dis}&=  \sum_{n=1,s}^{N_{s}} \delta \mu_n \, c_{n,s}^{\dagger} \, c_{n,s},
\end{align}
to the electronic Hamiltonian Eq. (\ref{eq:H_el}). Here, the parameters $\delta \mu_n$ are uniformly-distributed random variables within the range $\left[-W,W\right]$. Physically, this term can represent, for example, the presence of random impurities in the SiO$_2$ film or the imperfect planarity of the tPA molecule on top of the substrate.  In the presence of this term, Eqs. (\ref{eq:inversion}) and (\ref{eq:particle_hole}) are no longer valid, and therefore both inversion and particle-hole symmetries are lifted. According to the Cartan classification of topological phases \cite{Ryu10_Topological_classification, Kitaev_TI_classification}, in the absence of protecting symmeties, the SSH model is in a trivial topological phase. From the perspective of our work, this means that the topological quantization of the charges bound at the gates should be lost. In Fig. \ref{fig:disorder} we show the same results as in Fig. \ref{fig:energy_and_charge}(b), for a particular realization of the disorder random potential $\left\{\delta \mu_n \right\}$ and for different values of the amplitude parameter $W$, expressed in units of  $\Delta_0$. As expected,  as the the amplitude of the disorder potential increases, charge quantization is gradually lost. We nevertheless stress that the deviations from exact quantization are very small and  require a large amplitude of disorder potential. In practice, this means that charge quantization can be reasonably preserved by minimizing the presence of disorder and imperfections in the fabrication process. 

\section{Summary and Conclusions}\label{sec:conclusions}
We have theoretically studied a proposal for a polymer-based topological nano-device, where different topological phases and ground states can be externally controlled via locally applied gate voltages, $V_g$. Using an SSH-like model, we demonstrated the device's ability to accumulate topologically protected quantized charges in the regions near the gates. By implementing a self-consistent approach to numerically solve the set of non-linear equations coupling electronic and lattice degrees of freedom, we found robust solutions for the lattice configuration ${u_n}$ that display multiple topological kinks as $V_g$ is varied. Throughout this work, we have emphasized that our perfectly anti-symmetric and sufficiently distant gates should be understood as a simplified way to model the effects of a single gate acting on a tPA molecule, while minimizing theoretical complexity. The anti-symmetry of the gates is in fact a ficticious symmetry which can be justified due to the fact that we are focusing on the properties of a single gate. Grand-canonical ensemble calculations, which require the inclusion of an explicit coupling to a reservoir are currently underway, and are the subject of our subsequent line of research.

The local symmetry breaking induced by the gates, in an otherwise inversion-, particle-hole-, and time-reversal-symmetric model, is crucial to understand our results. This breaking allows us to explore different topological sectors within our modified SSH model. These phases are characterized by a $\mathbb{Z}$ topological invariant, represented either by the number of quantized charges, $Q_g = \pm ne$, bound at each gate, or by the number of kinks (per gate) in the ground-state lattice configurations. Due to their topological origin, these quantized charges and configurations are remarkably stable as functions of the external gate voltage $V_g$, except at specific points $V_g^{[n]}$, where topological transitions occur between solutions with $n-1$ and $n$ kinks. At these points, the electronic subgap spectrum of the device is fully reconstructed to accommodate the extra electronic charge. This phenomenological behavior contrasts with the topological phases typically found in other theoretical treatments of the SSH model, which, due to constraints imposed by local symmetries, are often described by an effective $\mathbb{Z}_2$ topological invariant. A key distinction in our case is that, because of the aforementioned local symmetry breaking, the electronic spectrum shown in Fig. \ref{fig:eigenenergies} is fully gapped, and no zero-energy states occur, as they do in the case of topological end-states commonly found in traditional SSH model studies \cite{Heeger88_Solitons_in_conducting_polymers,Su79_Solitons_in_polyacetylene,Su80_Soliton_excitations_in_PA,Asboth16_Short_course_on_TIs}.
In this context, we note that the existence of topological end states in our device is not relevant to the purposes of this work, which primarily focuses on the effect of gate voltages on the topological phase diagram. In fact, although we haven't explicitly discussed them here, \textit{we do observe} such localized end states. However, they are of no consequence for our objectives, as they are confined to a few sites at the ends of the chain. Moreover, they do not appear at zero energy due to the effect of the stretching term $\Gamma$ in the Hamiltonian, Eq. (\ref{eq:H_latt}).

Due to the topological origin of the induced charges, we believe our phenomenological model effectively captures the key characteristics of the voltage-controlled device. We further speculate that more detailed microscopic models are unlikely to qualitatively change our main conclusions, particularly concerning the electron-electron interactions, which have been omitted in this study. Given the energy gap between the ground state and the lowest-lying excited states (estimated to be approximately 24, 157, and 126 meV at the points of maximal stability in the first three topological sectors respectively), configurations with a small number of localized charges are expected to remain stable. Nevertheless, we intuitively speculate that the critical voltages $V^{[n]}$ should be pushed to larger values to compensate for the Coulomb repulsion term. Rather than being detrimental, this effect will stabilize the localized charges even further. The potential to induce strong local correlations could open new possibilities, such as the existence exotic ground states involving localized magnetic moments or the Kondo effect \cite{hewson}. Moreover, recent experiments have demonstrated superconductivity in 
$\alpha$-terphenyl polymers upon doping \cite{Zhong18_Superconductivity_in_p_terphenyl, Yan19_Superconductivity_in_p_quaterphenyl, Huang19_Superconductivity_in_p_terphenyl}, with strong correlations invoked to explain these intriguing results. Regarding the finite temperature effects, we  expect our results to be  robust since our estimated stabilization energies are larger than the thermal energy at room temperature, i.e., $T=300K\sim 26$ meV.

From a practical perspective, the proposed device is comparable to a semiconductor quantum dot (QD) in the Coulomb-blockade regime, which exhibits charge quantization due to its discrete energy spectrum and Coulomb interactions. However, unlike that system, the charge-quantization mechanism  is inherently topological and persists even in the absence of electron interactions. This key distinction, also reflected in the fact that the localized quantized charge cannot be associated with specific states in the spectrum (unlike in QDs),  underscores the system’s topological nature. Instead, charge quantization emerges here as a \textit{collective phenomenon} where all occupied states contribute. From a practical standpoint, the topological nature of these charges makes the device potentially more robust against disorder effects compared to QDs.

Concerning possible experimental realizations, we understand that  building single-molecule electronic devices presents a  challenge. However,  we believe that the device proposed in this work may be within experimental reach in the near future. Significant progress has been made in the on-surface synthesis of $\pi$-conjugated molecules in recent years \cite{Grill2007, Shen17_Frontiers_on_surface_review, Han21_Surface_assisted_fabrication_low_dimensional_carbon_nanostructures}. Today, single molecules can be observed and manipulated using advanced atomic-force microscopy (AFM) and scanning-tunneling microscopy (STM) techniques \cite{Wang19_Solitons_in_individual_PA_molecules}. Additionally, nanometric circuits can now be fabricated through electron-beam and photolithography \cite{Sharma22_Evolution_in_lithography_techniques}. Recent theoretical models also support the feasibility of designing single-molecule electronics \cite{Yao19_Unconventional_nanofabrication_supramolecular_electronics}, particularly with conductive polymers beyond tPA \cite{cirera2020tailoring}.

In conclusion, we believe that polymer-based integrated circuits, like the one described in this work, represent an exciting and experimentally feasible new platform where topological excitations can be externally controlled and manipulated to fabricate novel and more robust electronic nanodevices using current technology.

\begin{acknowledgments}
This work was supported by CONICET and Agencia
I + D + i under PICT 2017-2081, Argentina. A.M.L. is grateful to Pablo G. Roura for enlightening discussions. A.I.B. and L.M.A  thank CONICET for the doctoral fellowship
\end{acknowledgments}

\appendix

\section{Relation between the charge accumulated at the gates and the total energy}\label{sec:appendix}

In this section we will show that the contribution of the capacitive coupling term to the total energy is simply the electrostatic potential energy of the charges accumulated per gate due to the applied external voltage.

Our goal is to rewrite the total energy of the chain in terms of excess charges and external voltages. To this end, we start by writing the complete system Hamiltonian of Eq.~(\ref{eq:H}), making explicit only the electronic term that introduces the effect of the external gate voltages:
\begin{align}
\hat{H} &= \hat{H}^{0} + \sum_{n}^{N_{s}}\sum_{s} \mu_{n} \, c_{n,s}^{\dagger} \, c_{n,s},
\end{align}
For ease of interpretation, we begin by defining the operator~$\hat{q}_{n}$, which corresponds to the observable electronic population per site:
\begin{align}
\hat{q}_{n} &= \sum_{s} c_{n,s}^{\dagger} \, c_{n,s}. 
\end{align}
In terms of this quantity, the Hamiltonian of the system reads
\begin{align}
\hat{H} &= \hat{H}^{0} + \sum_{n}^{N_{s}} \mu_{n} \, \hat{q}_{n},
\end{align}
and the associated total energy expression can be expressed as follows:
\begin{align}
E_{\mathrm{GS}} &= \langle\, \hat{H} \,\rangle_{\mathrm{GS}} = \langle \, \Psi_{GS} \, | \, \hat{H} \, | \, \Psi_{GS} \, \rangle \\
E_{\mathrm{GS}} &= \langle\, \hat{H}^{0} \,\rangle_{\mathrm{GS}} + \sum_{n}^{N_{s}} \mu_{n} \, q_{n},
\end{align}
where $q_{n} = \left\langle \, \hat{q}_{n} \, \right\rangle$ is the average of the charge at site $n$.  The above energy expression allows us to more easily make explicit the excess charge per site, $\delta q_{n}$, which can be interpreted as the net charge per site (${\delta q_{n} = q^{\text{ion}}_{n} - q_{n}}$) or as the charge resulting from the fluctuation of the electron density at each site with respect to the neutral reference in the absence of external voltages (${\delta q_{n} = -(q_{n} - q_{n}^{0})}$), since both interpretations give identical expressions (as ${q^{\text{ion}}_{n} = q_{n}^{0} = 1}$):
\[
\delta q_{n} = 1 - q_{n},
\]
Note that the sign convention of this definition of charge allows a direct physical interpretation: excess electron density results in a net negative charge, and electron depletion results in a net positive charge.

We then continue working on our expression for the total energy by replacing with (i) our definition of excess charge per site and (ii) with the values of $\mu_{n}$ according to our model of two anti-symmetric gate voltages (see Fig.~1 in the main manuscript):
\begin{align}
E_{\mathrm{GS}} &= \langle\, \hat{H}^{0} \,\rangle_{\mathrm{GS}} + \sum_{n}^{N_{s}} \mu_{n} \, (1 - \delta q_{n}),\\
E_{\mathrm{GS}} &= \langle\, \hat{H}^{0} \,\rangle_{\mathrm{GS}} + \sum_{n \in g^{+}}^{w_{g}} (+V_{g}) \, (1 - \delta q_{n})\nonumber  \\
&+ \sum_{n \in g^{-}}^{w_{g}} (-V_{g}) \, (1 - \delta q_{n}),
\end{align}
which simplifies to:
\begin{align}
E_{\mathrm{GS}} &= \langle\, \hat{H}^{0} \,\rangle_{\mathrm{GS}} - \sum_{n \in g^{+}}^{w_{g}} V_{g} \, \delta q_{n} - \sum_{n \in g^{-}}^{w_{g}} (-V_{g}) \, \delta q_{n}.
\end{align}
Finally, after recognizing the charge accumulated at the gate (${Q_{g} \ge 0}$), we can conclude that the capacitive coupling term contributes to the total energy with the electrostatic potential energy of these excess charges per gate due to the applied external voltage:
\begin{align}
Q_{g} &= \pm \sum_{n \in g^{\pm}}^{w_{g}} \delta q_{n},\\
E_{\mathrm{GS}} &= \langle\, \hat{H}^{0} \,\rangle_{\mathrm{GS}} - 2\, V_{g} \, Q_{g}.
\end{align}
The factor $2$ appears because our model works with two anti-symmetric gate voltages (i.e., of equal width, opposite applied voltage, and equal distance from the center of the chain).

%
\end{document}